\documentclass[prl,superscriptaddress,floatfix,amsmath,notitlepage,twocolumn]{revtex4-1}
\usepackage{graphicx}
\usepackage{color}
\usepackage{siunitx}
\usepackage[normal]{subfigure}
\usepackage[usenames,dvipsnames]{xcolor}
\usepackage{amssymb}
\usepackage{dsfont}
\newcommand{\be}{\begin{equation}}
\newcommand{\ee}{\end{equation}}

\usepackage[colorlinks=true,urlcolor=blue,linkcolor=blue,citecolor=blue]{hyperref}

\begin{document}

\title{Chiral Light--Matter Interaction Beyond the Rotating-Wave Approximation}

\author{Sahand Mahmoodian}
\affiliation{Institute for Theoretical Physics, Institute for Gravitational Physics (Albert Einstein Institute), Leibniz University Hannover, Appelstra{\ss}e 2, 30167 Hannover, Germany}

\date{\today}

\begin{abstract}
I introduce and analyse chiral light--matter interaction in the ultrastrong coupling limit where the rotating-wave approximation cannot be made. Within this limit, a two-level system (TLS) with a circularly polarized transition dipole interacts with a copolarized mode through rotating-wave terms. However, the counter-rotating terms allow the TLS to couple to a counter-polarized mode with the same coupling strength, i.e., one that is completely decoupled within the rotating-wave approximation. Although such a Hamiltonian is not particle number conserving, the conservation of angular momentum generates a $U(1)$ symmetry which allows constructing an ansatz. The eigenstates and dynamics of this novel model are computed for single-cavity interactions and for a many-mode system.  The form of the ansatz provides significant analytic insight into the physics of the ground state and the dynamics, e.g., it indicates that the ground states are two-mode squeezed. This work has significant implications for engineering light--matter interaction and novel quantum many-body dynamics beyond the rotating-wave approximation.  
\end{abstract}

\maketitle

The Rabi model \cite{Rabi1937PR} constitutes perhaps the paradigmatic example of light--matter interaction (LMI) in quantum optics. It describes the interaction of a two-level system (TLS) with a single cavity mode under the dipole approximation. 
Recently, this model has has seen renewed interest for two main reasons: for many years LMI was considered under the rotating-wave approximation (RWA) where terms that do not preserve the total number of excitations are neglected. This is typically valid for systems operating at optical frequencies. Recently, however, microwave-frequency circuit QED platforms with  light--matter coupling $g/\omega_c \sim 1$ have been developed \cite{Niemczyk2010NPHYS, FornDiaz2017NPHYS, Yoshihara2017NPHYS, Kockum2019NatRevPhys}. These platforms can thus probe the full dynamics of the Rabi model. The second reason is that recent work by Braak \cite{Braak2011PRL} has piqued theoretical interest by showing that the Rabi model is analytically solvable: a task which remained elusive for many years. This has led to many works investigating the structure of the solution of the Rabi model and its extensions \cite{Moroz2014AP, Tomka2014PRA, Xie2014PRX}.

Simultaneously, tremendous effort has been put in engineering LMI. This has involved using photonic nanostructures to strengthen coupling to a single mode while minimizing coupling to others \cite{Arcari2014PRL, Goban2014NCOM, Lodahl2015RMP}. Researchers have now also developed sophisticated techniques that allow tailoring the phase and magnitude of the coupling to each mode. For example, this has been achieved using non-local interactions in one- \cite{Kockum2018PRL} and two-dimensional systems \cite{GonzalezTudela2019arXiv}, by considering emission in the presence of strong dispersion \cite{GonzalezTudela2017PRL},  incorporating phonons \cite{Calajo2019arXiv}, as well by using chiral LMI \cite{Petersen2014Science, Mitsch2014NCOM, Sollner2015NNANO, Lodahl2017Nature}.  Chiral LMI uses the circularly polarized transition dipole of a TLS to selectively couple to modes whose electric fields have the same circular polarization at the position of the TLS, while remaining decoupled to counter-circulating modes. Unidirectional emission from a TLS is achieved by engineering the electromagnetic modes of a one-dimensional waveguide to have a direction-dependent circular polarization \cite{Bliokh2012PRA, Mahmoodian2017OME}. This then enables emission whose direction is controlled by the handedness of the TLS's transition dipole. Chiral LMI has however thus far only been considered within the RWA.

In this manuscript, I show that chiral LMI also leads to novel physics beyond the RWA and can be used to engineer interactions in the ultrastrong-coupling limit. In particular, I show that, when a TLS with a circularly polarized transition dipole couples to a bath whose modes are elliptically or circularly polarized, the coupling strengths are generally different for the rotating-wave and counter-rotating-wave terms. Extraordinarily, a mode that is completely orthogonal to the TLS's transition dipole and does not couple within the RWA can interact with the TLS through the counter-rotating terms. I highlight this using a novel two-mode \textit{chiral Rabi model}. I then extend ultrastrong chiral LMI to a many-mode model and compute its ground states and its quench dynamics. In general, the physics of these systems can be described in terms of the conservation of angular momentum, which generates a $U(1)$ symmetry of the Hamiltonian. The conserved quantity is exploited to construct an ansatz for obtaining the ground states and dynamics in these systems. This work paves the way for engineering many-body dynamics in quantum optical systems beyond the RWA.

I begin by considering a TLS interacting with a single cavity mode under the dipole approximation. This has the interaction Hamiltonian $\hat{H}_{\rm int} = -\hat{\mathbf d} \cdot \hat{\mathbf E}_a$, where the electric field operator for mode $a$ is $\hat{\mathbf E}_a=\mathbf E_a \hat{a} + \mathbf E_a^* \hat{a}^\dagger$ with cavity field $\mathbf E_a$, where $\hat{a}$ ($\hat{a}^\dagger$) is an annihilation (creation) operator. The dipole operator is  $\hat{\mathbf d} = \mathbf d \hat{\sigma}_- + \mathbf d^* \hat{\sigma}_+$ where $\mathbf d$ is the transition dipole moment and $\hat{\sigma}_-=|g\rangle \langle e|$ and  $\hat{\sigma}_+=|e\rangle \langle g|$ take the TLS to and from the excited $| e\rangle$ and ground $|g\rangle$ states.  The interaction Hamiltonian then clearly has the form $\hat{H}_{\rm int} = g_{R} \hat{\sigma}_- \hat{a}^\dagger + g_{R}^* \hat{\sigma}_+ \hat{a} + g_{cR} \hat{\sigma}_- \hat{a} + g_{cR}^* \hat{\sigma}_+ \hat{a}^\dagger$, where $g_{R} = \mathbf d \cdot \mathbf E_a^*$ and $g_{cR} = \mathbf d \cdot \mathbf E_a$. For a linearly polarized transition dipole and a linearly polarized dipole $g_{R} = g_{cR}$, but when the transition dipole or the electric field are elliptically polarized, i.e. when $\mathbf d$ or $\mathbf E$ cannot be made real, $g_{R} \neq g_{cR}$. Notably, the interaction does not generally take on the form of the Rabi model and cannot be written as $(\sigma_- + \sigma_+) (\hat{a} + \hat{a}^\dagger)$. Instead it has the form of the generalized Rabi model \cite{Xie2014PRX} where the rotating wave (RW) and counter-rotating wave (cRW) terms have different coupling coefficients. Previous work \cite{Xie2014PRX} suggested creating a generalized Rabi model using both electric and magnetic dipole moments. However, simply controlling the degree of circular polarization of the electric dipole and cavity modes enables engineering the relative strength of the RW and cRW parts of the Hamiltonian. Engineering a Hamiltonian within the generalized Rabi model is highly desirable as it can, e.g., be used to simulate supersymmetric quantum field theories \cite{Tomka2015SciRep} and electron transport in the presence of spin--orbit coupling \cite{Erlingsson2010PRB}.


%
\begin{figure}[!t]
\includegraphics[width=\columnwidth]{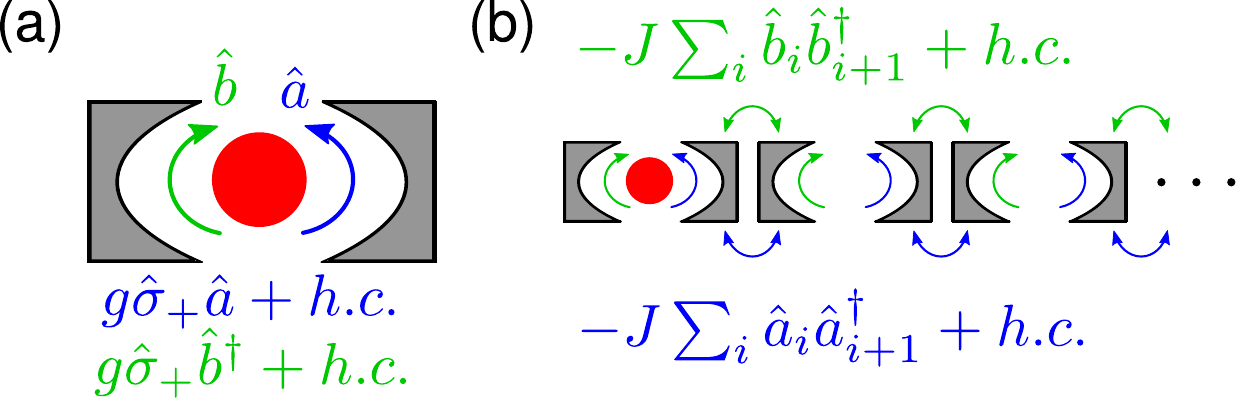}
\caption{\label{fig:schem} (a) Schematic of the single-cavity chiral Rabi model. A two-level system with a circularly-polarized dipole moment (red circle) interacts with the copolarized mode $a$ through rotating wave terms (blue) and with a counter-polarized mode through counter-rotating terms (green). (b) A many-mode open system is composed by introducing an array of cavities with nearest-neighbour coupling. The $a$ and $b$ modes are orthogonal and only couple through the two-level system.}
\end{figure}

Conservation of angular momentum underlies the difference between the values of $g_{cR}$ and $g_{R}$ when the fields are not linearly polarized. In order to fully take advantage of this,  I now consider the limit where both $\mathbf E_a$ and $\mathbf d$ are circularly polarized and have the same polarization. In this limit the counter-rotating terms vanish and the interaction is purely through the rotating-wave terms. The role of angular momentum becomes clear by considering a cavity which also supports another mode with field $\mathbf E_b$ and annihilation operator $\hat b$ which has the opposite circular polarization to $\mathbf E_a$ and $\mathbf d$, i.e. $\mathbf E_b = \mathbf E^*_a$ (see Fig.~\ref{fig:schem}(a)). This can occur if the cavity has a point group symmetry such that it supports two degenerate polarization modes \cite{McIsaac1975IEEE, Steel2001OL}. In this manuscript, I, for simplicity, assume that the two cavity modes are spectrally degenerate with frequency $\omega_c$, but this is not generally a requirement. The total Hamiltonian ($\hbar = 1$) under the dipole approximation is
\begin{equation}
\label{eq:Hamiltonian}
\hat{H} = \frac{\omega_0}{2}\hat{\sigma}_z + \omega_c (\hat{a}^\dagger \hat{a} + \hat{b}^\dagger \hat{b}) + g \, \hat{\sigma}_+ (\hat{a} + \hat{b}^\dagger ) + h.c.,
\end{equation}
where $h.c.$ denotes the Hermitian conjugate of the preceding term, and $g=\mathbf d \cdot \mathbf E^*_a = \mathbf d \cdot \mathbf E_b$. Here, $\hat{\sigma}_z$ is the Pauli $z$-matrix and $\omega_0$ is the transition frequency of the TLS. In this \textit{chiral Rabi model} the TLS interacts with the copolarized mode through the rotating wave terms, while it interacts with the orthogonally polarized mode through counter-rotating terms. The novel model highlights that modes whose fields are orthogonal to the transition dipole moment, and therefore do not interact in the rotating wave approximation, couple through counter-rotating terms with the same coupling coefficient. This is a consequence of the conservation of angular momentum: rotating-wave terms transfer excitations between the TLS and cavity mode, thus the transition dipole and field must have the same angular momentum. On the other hand, counter-rotating terms create or destroy excitations in pairs, thus the dipole moment and field must have opposite angular momenta. The angular momentum of the system in the $z$-direction, 
\begin{equation}
\hat{L}_z = \hat{a}^\dagger \hat{a} - \hat{b}^\dagger \hat{b} + \frac{\hat{\sigma}_z}{2},
\end{equation}
is conserved and commutes with the Hamiltonian $[\hat{L}_z,\hat{H}]=0$. It generates a continuous $U(1)$ symmetry of the system. In the Jaynes-Cummings model the total number of excitations and the angular momentum are conserved. On the other hand, the conserved quantity  $\hat{L}_z$  does not correspond to the total number of excitations which is not conserved. This type of conserved quantity has previously been observed in other quantum many-body systems \cite{Porras2012arXiv, Fan2014PRA}.

\begin{figure}[!t]
\includegraphics[width=\columnwidth]{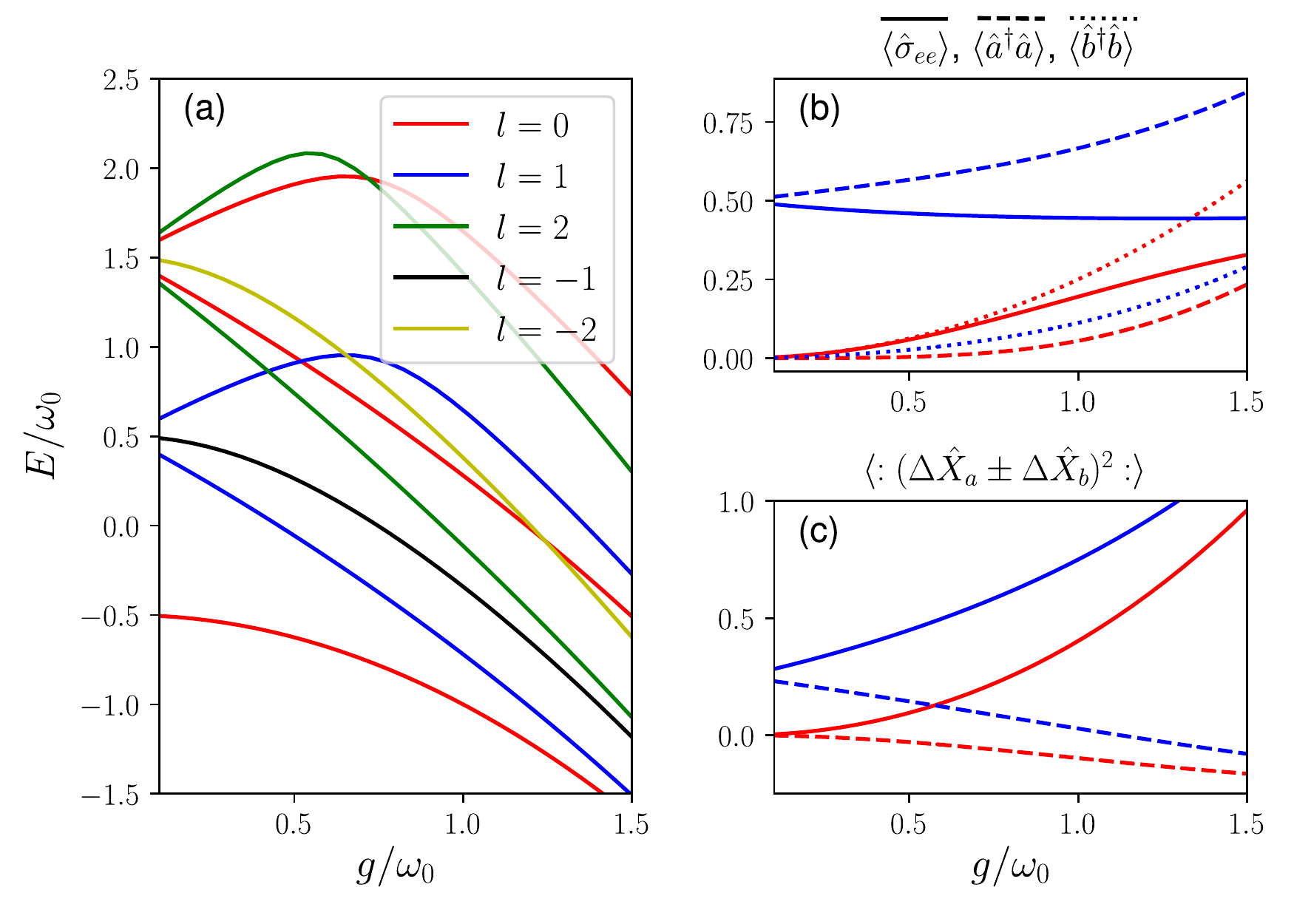}
\caption{\label{fig:eigs} (a) First several eigenenergies of the chiral Rabi Hamiltonian for different angular momentum quantum numbers $l$ for $\omega_0=\omega_c$. (b) Observables of the lowest energy eigenstates in the  $l=0$ (red) and $l=1$ (blue) manifolds. The plot shows the excited-state population $\langle \hat{\sigma}_{ee} \rangle$ (solid lines), $\langle \hat{a}^\dagger \hat{a} \rangle$ (dashed lines), and $\langle \hat{b}^\dagger \hat{b} \rangle$ (dotted lines). (c) Normally ordered variance $\langle : ( \Delta \hat{X}_a + \Delta \hat{X}_b)^2 :\rangle$ (solid lines) and $\langle : (  \Delta \hat{X}_a - \Delta \hat{X}_b)^2 :\rangle$ (dashed lines). Squeezing occurs for values below zero.}
\end{figure}

The eigenstates of the conserved quantity $\hat{L}_z$ can be used to construct the eigenstates of $\hat{H}$ using the ansatz
\begin{equation}
\label{eq:ansatz}
| \psi \rangle_l = \sum_{n=0}^\infty c_{n,l}|g \rangle |n+l,n\rangle + \sum_{n=1}^\infty  d_{n,l}| e \rangle |n+l-1,n\rangle,
\end{equation}
where $|i,j \rangle$ indicates $i$-photon and $j$-photon Fock states occupying modes $a$ and $b$ respectively. Substituting the ansatz in Schr\"{o}dinger's equation leads to a set of eigenrecurrence relations for the coefficients $c_{n,l}$ and $d_{n,l}$ and the energy $E_l$. These equations appear in the Supplementary Material (SM). The system of recurrence relations is diagonalized by truncating the Hilbert space at a sufficiently large value of $n$. Figure \ref{fig:eigs}(a) shows the eigenstates for different values of $l$ versus $g$. When $g \ll \omega_0$ the eigenstates are Jaynes-Cummings-like, and when $\omega_0 = \omega_c$, $|\psi \rangle^\pm_{\rm JC} = \frac{1}{\sqrt{2}} \left[ |g \rangle | n, m \rangle \pm |e \rangle | n-1, m \rangle \right]$ with energy $E_{\rm JC}^{\pm} = (n+m-\frac 1 2)\omega_0 \pm g \sqrt{n}$ for integers $n \ge 0 $ and $m \ge 0$. Here the photons in modes $a$ and $b$ are uncorrelated and the ground state is trivial. As $g$ increases the $l=0$ ground state becomes composed of an entangled state of light and matter with photons in modes $a$ and $b$ and the TLS partially excited (see Fig.~\ref{fig:eigs}(b)). The entanglement entropy between the TLS and the cavity modes approaches unity as $g/\omega_0$ increases (see SM). I note that, although a TLS has been considered here, the treatment can be generalized to other level schemes (see SM for a $V$-level scheme).

The form of the ansatz (\ref{eq:ansatz}) reveals how the TLS and the photonic modes are correlated. The form of the Fock states is reminiscent of two-mode squeezing. This is considered by introducing quadratures $\hat{X}_a = (\hat{a} + \hat{a}^\dagger)/2$ and $\hat{P}_a = i(\hat{a}^\dagger - \hat{a})/2$ (with equivalent definitions for mode $b$). For all eigenstates $\langle \hat{a} \rangle = \langle \hat{b} \rangle = \langle \hat{a}^2 \rangle = \langle \hat{b}^2 \rangle = \langle \hat{a}^\dagger \hat{b} \rangle=0$. From this one can compute that the normally ordered variances for the individual quadratures are $\langle : (\Delta \hat{X}_a)^2 : \rangle = \langle : (\Delta \hat{P}_a)^2 : \rangle = \langle \hat{a}^\dagger \hat{a} \rangle /2$ and $\langle : (\Delta \hat{X}_b)^2 : \rangle = \langle : (\Delta \hat{P}_b)^2 : \rangle = \langle \hat{b}^\dagger \hat{b} \rangle /2$, which are always positive and therefore the individual mode quadratures are not squeezed. Additionally both modes also satisfy $ \langle : \Delta \hat{X} \Delta \hat{P} : \rangle=0$. On the other hand, there are strong correlations between quadratures of the different modes with  $\langle : \Delta \hat{X}_a \Delta \hat{X}_b :\rangle = -\langle : \Delta \hat{P}_a \Delta \hat{P}_b :\rangle = \operatorname{Re}{\langle \hat{a} \hat{b} \rangle} /2$. Figure~\ref{fig:eigs}(c) shows the variance $\langle : (\Delta \hat{X}_a \pm  \Delta \hat{X}_b)^2 :\rangle = \langle : (\Delta \hat{P}_a \mp \Delta \hat{P}_b)^2 :\rangle$  for the ground state $l=0$ and the lowest energy eigenstate of $l=1$. The $l=0$ ground state exhibits two-mode squeezing for all $g$ while the $l=1$ state is squeezed for $g \gtrsim \omega_0$ and the strength of the squeezing grows with $g$. 

Using the ansatz, the dynamics of the chiral Rabi model can also be computed. In the large $g$ limit, the counter-rotating terms destroy the coherence of the Rabi oscillations and the system descends into quasiperiodic collapse and revivals similar to those found in the Rabi model \cite{Casanova2010PRL} (see SM).

\begin{figure}[!t]
\includegraphics[width=\columnwidth]{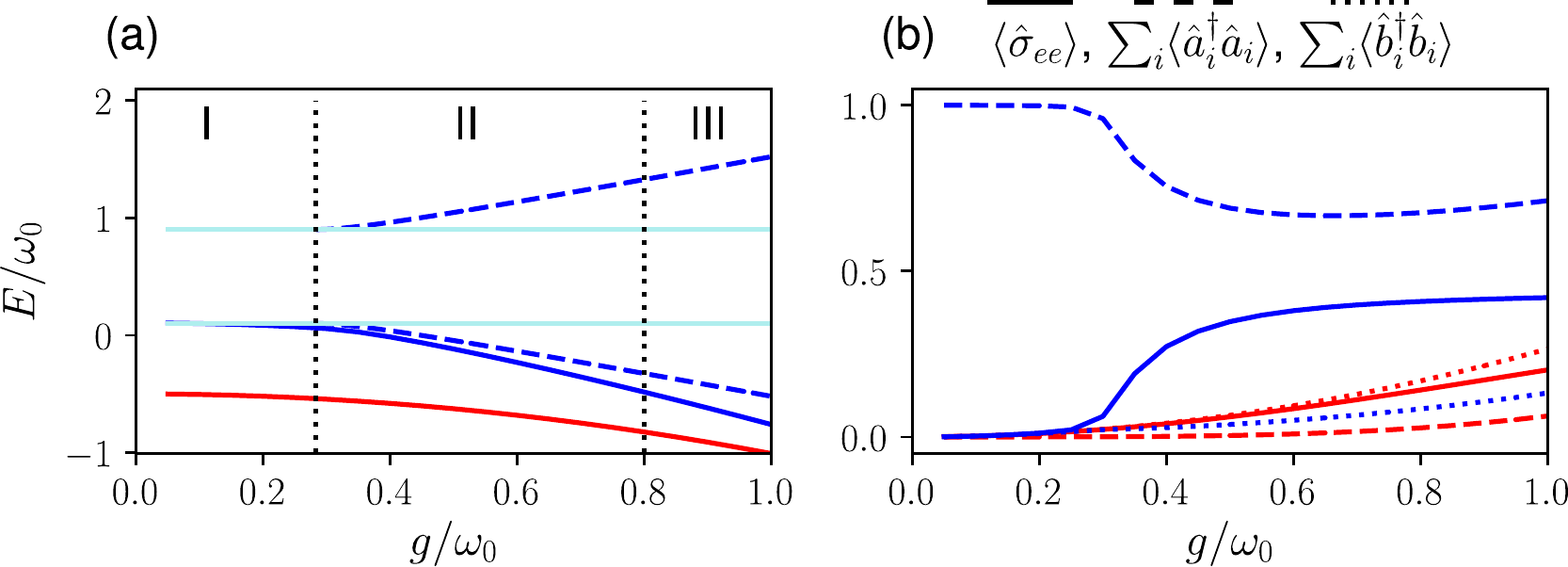}
\caption{\label{fig:MBGS} (a) Ground state energy of the many-body Hamiltonian for the $l=0$ (red) and $l=1$ (blue) manifolds versus coupling coefficient $g$ computed by truncating the ansatz at $n=2$ with $L=20$ sites. Here, $J=0.2 \omega_0$ and $\omega_0=\omega_c$. The dashed blue line shows the single excitation bound states computed within the RWA and the region within the horizontal cyan lines is the photon band. The symbols I, II, III show the different phases of the $l=1$ subspace (see main text). (b) Observables for the $l=0$ (red) and $l=1$ (blue) ground states.}
\end{figure} 

The idealized single-cavity model successfully illustrates the role of conservation of angular momentum in chiral LMI beyond the RWA. Nevertheless, most quantum optical platforms exhibit open-system dynamics where dissipation plays a key role in the system evolution.  This has recently been demonstrated beyond the RWA in circuit QED platforms \cite{FornDiaz2017NPHYS} and has been supported by numerical and analytic investigations \cite{SanchezBurillo2014PRL, Shi2018PRL}. I now consider chiral LMI beyond the RWA in a one-dimensional many-mode model. In chiral LMI \cite{Petersen2014Science, Mitsch2014NCOM, Sollner2015NNANO, Lodahl2017Nature}, polarization selection rules allow the TLS to couple to a unidirectional spatial mode. Here, instead of considering decoupled directional modes, I consider a photonic bath formed by an array of $L$ coupled cavities, each with two polarization modes. The two polarization modes $a$ and $b$ propagate independently in analogy to the forward and backward propagating modes in conventional chiral LMI (see Fig.~\ref{fig:schem}(b)). The full Hamiltonian for this system is      
\begin{equation}
\label{eq:MBHamiltonian}
\begin{split}
\hat{{\cal H}} &= \frac{\omega_0}{2}\hat{\sigma}_z +\omega_c \sum_{i=0}^{L-1} \hat{a}_i^\dagger \hat{a}_i + \hat{b}_i^\dagger \hat{b}_i + g \, \hat{\sigma}_+ (\hat{a}_0 + \hat{b}_0^\dagger ) + h.c. \\
&   - J \sum_{i=0}^{L-2} \left(\hat{a}_i^\dagger \hat{a}_{i+1} + \hat{b}_i^\dagger \hat{b}_{i+1} \right) + h.c.,
\end{split}
\end{equation}
where $\hat{a}_i$ and $\hat{b}_i$ are the annihilation operators for the $a$ and $b$ modes of the $i$th cavity, and $J$ is the cavity coupling coefficient. Here the coupled-cavity array  provides a simple model for a bath whose dispersion curve can be computed exactly with each set of modes having the dispersion $\omega_{a/b}(k) = \omega_c - 2J \cos{(k)}$, where $k \in [-\pi, \pi)$. The angular momentum operator is now $\hat{L}_z = \hat{\sigma}_z/2 + \sum_i \hat{a}_i^\dagger \hat{a}_i - \hat{b}_i^\dagger \hat{b}_i$ and commutes with $\hat{{\cal H}}$. As before, the eigenstates with the same angular momentum can be used to construct an ansatz
\begin{equation}
\label{eq:MBansatz}
\begin{split}
&| \Psi \rangle_l = \sum_{n=0}^\infty\sum_{\mathbf i, \mathbf j=0}^{L-1} \frac{c_{n,l;j_1,j_2, \ldots, j_{n}}^{ i_1,i_2, \ldots, i_{n+l}}}{\sqrt{n! (n+l)!}} \prod_{k,m=1}^{n+l,n} \hat{a}^\dagger_{i_{k}}  \hat{b}^\dagger_{j_{m}}  |0 \rangle |g \rangle\\
& +  \frac{d_{n,l;j_1,j_2, \ldots, i_{n}}^{ i_1,i_2, \ldots, i_{n+l-1}}}{\sqrt{n! (n+l-1)!}} \prod_{k,m=1}^{n+l-1,n} \hat{a}^\dagger_{i_{k}}  \hat{b}^\dagger_{j_{m}} |0 \rangle |e \rangle.
\end{split}
\end{equation}
\begin{figure}[!t]
\includegraphics[width=\columnwidth]{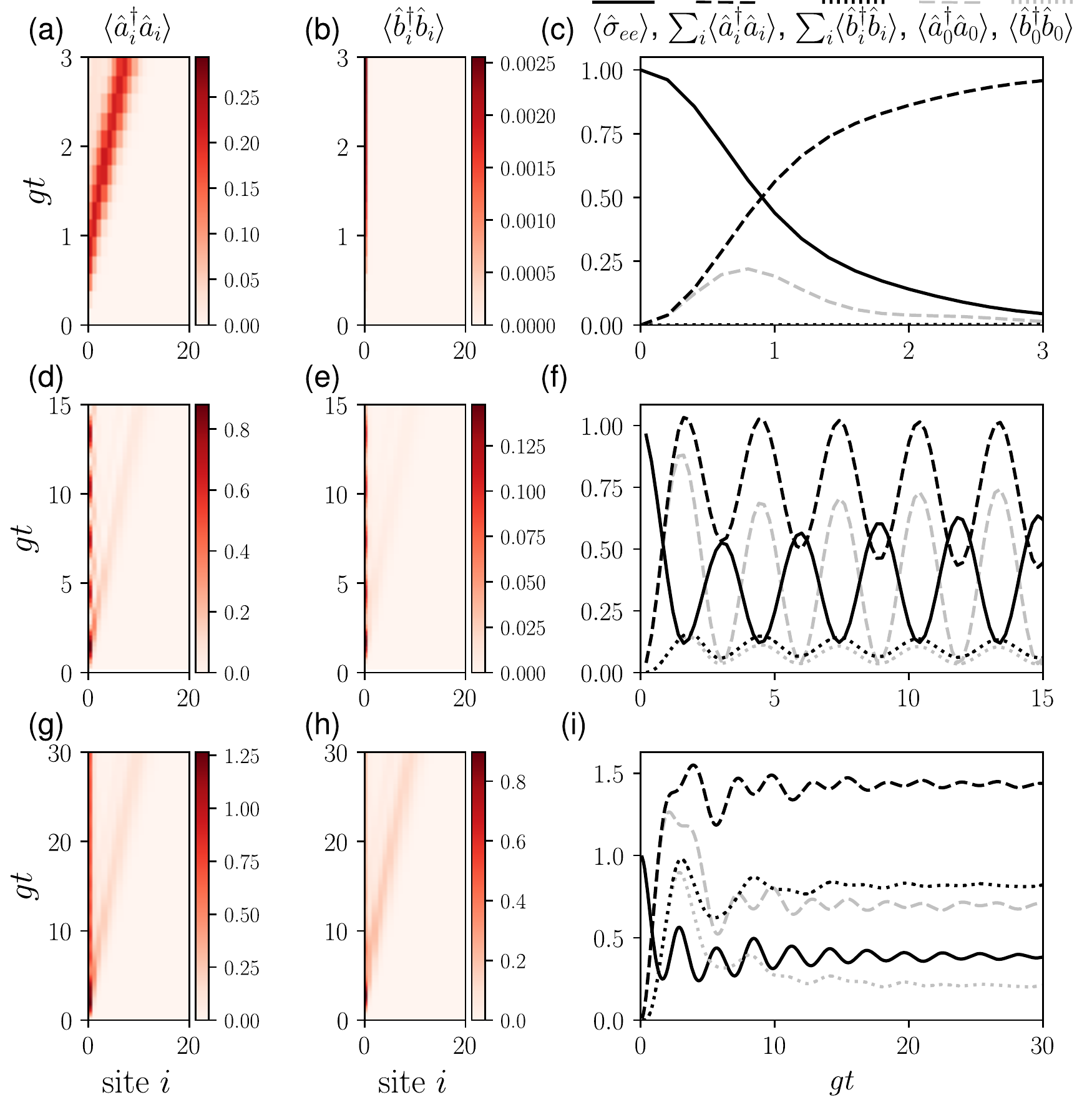}
\caption{\label{fig:MBDynamics} Many-body time dynamics starting in the $| e \rangle | 0 0 \rangle$ state for $\omega_0=\omega_c$ and $J=0.2\omega_0$ with (a)-(c) $g=0.1\omega_0$, (d)-(f) $g=0.5\omega_0$, and (g)-(i) $g=\omega_0$, computed using MPS (see SM for details). Left (center) column shows number of photons in the $a$ ($b$) mode versus site index $i$ and normalized time $g t$. The right column shows observables versus time: the excited state population of the TLS $\langle \hat{\sigma}_{ee}\rangle$ (solid black), photons in the $a$ mode $\sum_i \langle \hat{a}_i^\dagger \hat{a}_i \rangle$ (dashed black), in the $b$ mode $\sum_i \langle \hat{b}_i^\dagger \hat{b}_i \rangle$ (dotted black), and the number of photons in cavity $i=0$ for mode $a$ $\langle \hat{a}_0^\dagger \hat{a}_0 \rangle$ (solid grey) and mode $b$ $\langle \hat{b}_0^\dagger \hat{b}_0 \rangle$ (dotted grey).}
\end{figure} 


In general, solving for the ground state or the dynamics of the Hamiltonian is a many-body problem with an exponentially large Hilbert space. Here the ground state and dynamics of (\ref{eq:MBHamiltonian}) are computed by truncating the ansatz (\ref{eq:MBansatz}) and by using a matrix-product state (MPS) ansatz \cite{Schollwock2011Annals}. Truncating the sum in (\ref{eq:MBansatz}) to $n \leq 2$, provides a good approximation for the ground states. The accuracy of the ansatz is limited by the number of photons that can be present in the system. Figure~\ref{fig:MBGS}(a) shows the energy of the ground state of the $l=0$ and $l=1$ manifolds. These results agree well with MPS calculations (see SM). 
The regions I, II, III show three different phases of the $l=1$ eigenstates. Region I has $g<\sqrt{2} J$ and there is no localized state outside the photon band. This is due to the semi-infinite nature of the lattice and also occurs within the RWA where the bound states can be computed analytically (see SM). This is unlike an infinite system where there is always a bound state \cite{Calajo2016PRA, Shi2016PRX, SanchezBurillo2017PRA}. The change from phase I to II is clearly visible in the observables in Fig.~\ref{fig:MBGS}(b).

In regions II and III the  $l=1$ ground state is composed of a photon-atom bound state whose energy lies below the photon continuum. Here the counter-rotating terms cause the energy of the eigenstate to decrease. Note that within the RWA there is also an eigenstate above the photon continuum. The numerical calculations were unable to resolve this upper eigenstate of the full Hamiltonian (\ref{eq:MBHamiltonian}). In the limit where $g \gg J$ the behaviour of the eigenstates approaches that of the single cavity eigenstate. Within this limit the upper eigenstate behaves like the upper $l=1$ eigenstate shown in Fig.~\ref{fig:eigs}(a), i.e., due to the counter-rotating terms its energy decreases as $g$ increases. This causes the eigenstate to eventually enter the photon continuum and it is no longer bound. This sets the boundary between phases II and III. From the single cavity eigenstate this point can be estimated to occur at $g \sim 0.8 \omega_0$.



One of the main features of chiral LMI under the RWA is the selective spontaneous emission of light into a directional subset of modes \cite{Sollner2015NNANO}. Computing the non-equilibrium dynamics of the Hamiltonian (\ref{eq:MBHamiltonian}) shows how the counter-rotating wave terms modify this behaviour. Here, the dynamics are computed using an open source MPS implementation \cite{Wall2012NJP, Jaschke2018CompPhysComm}. The MPS calculations are checked for convergence by varying the bond dimensions and the maximum number of bosons at each site (see SM). Figure \ref{fig:MBDynamics} shows the evolution of the system when starting in the state $|e \rangle | 0 0 \rangle$ for (a)-(c) $g=0.1\omega_0$, (d)-(f) $g=0.5\omega_0$, (g)-(i) $g=\omega_0$. These correspond to values lying in phases I, II, and II of Fig.~\ref{fig:MBGS}(a). As $g$ increases there are two key changes in the system dynamics: the population of the $b$ modes increases, and the dynamics undergoes changes from decay, to Rabi oscillations, and then to fractional decay into a bound state. The increase in photon population of the $b$ modes is explicitly due to the counter-rotating terms whose role becomes more prevalent as $g$ increases. This is the hallmark of the chiral Rabi model. 

The change in the nature of the dynamics is due to a combination of the nonlinear dispersion of the photon band and the counter-rotating interaction. It can be understood from the three phases of the $l=1$ ground state shown in Fig.~\ref{fig:MBGS}. In Fig.~\ref{fig:MBDynamics}(a)-(c) there is no bound state and the excited TLS can only decay into the photon continuum. Figure \ref{fig:MBDynamics}(d)-(f) corresponds to  region II which contains two bound eigenstates. The excited TLS emits into these two bound states which continue to beat together in time generating Rabi oscillations. As $g$ increases the Rabi oscillations are modified due to the counter-rotating terms and the $b$ mode is populated. In Fig.~\ref{fig:MBDynamics}(g)-(i) the dynamics exhibit fractional decay into a bound state. This occurs in region III. Since there is only a single bound state in his region, the dynamics cannot exhibit Rabi oscillations. Comparing observables in Figs.~\ref{fig:MBGS}(b) and \ref{fig:MBDynamics}(i) shows that the system decays into the $l=1$ ground state. The overlap of the states is $| \langle  e |\langle 0 0 | \Psi \rangle_{l=1}^{\rm ground} |^2=0.32$.

In conclusion, I have shown that chiral LMI forms a novel platform for exploring many-body physics beyond the RWA. In this limit, counter-circulating modes that are decoupled within the RWA play a key role in the system dynamics. The single-cavity chiral Rabi model can be potentially implemented in circuit QED platforms or using trapped cold atoms \cite{Dareau2018PRL} or ions \cite{Meekhof1996PRL}. A circuit QED implementation requires introducing angular momentum or chirality into the system, which was recently illustrated using three qubits \cite{Roushan2017NPHYS}. Many-mode chiral LMI can, in principle, be realized in circuit QED by coupling a qubit to two points in a transmission line with different phases \cite{Vermersch2016PRA}.  In general, the work here opens significant new avenues in the study of spin-Boson and Kondo physics \cite{Leggett1987RMP} with engineered impurity--bath interactions.

I would like to thank Klemens Hammerer for useful discussions and for proof reading this manuscript. I also acknowledge useful discussions with Anders S{\o}rensen, Florentin Reiter, Philipp Schneeweiss, and Tao Shi.

\bibliography{bigBib}

\onecolumngrid
\newpage

\pagebreak
\widetext
\begin{center}
\textbf{\large Supplementary Material: Chiral Light--Matter Interaction Beyond the Rotating-Wave Approximation}
\end{center}
\setcounter{equation}{0}
\setcounter{figure}{0}
\setcounter{table}{0}
\setcounter{page}{1}
\makeatletter
\renewcommand{\theequation}{S\arabic{equation}}
\renewcommand{\thefigure}{S\arabic{figure}}
\renewcommand{\thesection}{S\arabic{section}}

\section{Eigenrecurrence equation and single-cavity dynamics}
 
In this section I present the eigenrecurrence relations for the single-cavity chiral Rabi model. I start with the single-cavity Hamiltonian given in Eq.~\ref{eq:Hamiltonian} of the main text
\begin{equation*}
\hat{H} = \frac{\omega_0}{2}\hat{\sigma}_z + \omega_c (\hat{a}^\dagger \hat{a} + \hat{b}^\dagger \hat{b}) + g \, \hat{\sigma}_+ (\hat{a} + \hat{b}^\dagger ) + h.c.,
\end{equation*}
and the ansatz in Eq.~\ref{eq:ansatz}
\begin{equation*}
| \psi \rangle_l = \sum_{n=0}^\infty c_{n,l}|g \rangle |n+l,n\rangle + \sum_{n=1}^\infty  d_{n,l}| e \rangle |n+l-1,n\rangle.
\end{equation*}
Substituting these into Schr\"{o}dinger's equation and using the orthogonality of the Fock states, it is straightforward to obtain the set of coupled eigenrecurrence relations
\begin{equation}
\begin{split}
\label{eq:eigenRecurrence}
g \sqrt{n+1} \, d_{n+1,l} + g \sqrt{n+l} d_{n,l} + \left[\omega_c (2n+l) - \frac{\omega_0}{2} \right]c_{n,l} &= E_l c_{n,l} \\
g \sqrt{n+l} \, c_{n,l} + g \sqrt{n} \, c_{n-1,l} + \left[\omega_c (2n+l-1) + \frac{\omega_0}{2} \right]d_{n,l} &= E_l d_{n,l}.
\end{split}
\end{equation}
The coefficients $c_{n,l}$ and $d_{n,l}$ can be obtained by writing \ref{eq:eigenRecurrence} as a matrix eigenvalue equation and truncating at a sufficiently large Fock state. Note that when $l<0$ all Fock state coefficients with $n<l$ are zero.

Since one has to only solve for two coefficients for each Fock state the size of the problem scales linearly with the number of Fock states used. This means that it is feasible to compute a near-complete basis of eigenstates. These can then be used to compute the evolution of an arbitrary initial state. For example, given an arbitrary initial state $|\psi(0) \rangle$ the time evolution is given by
\begin{equation}
| \psi(t) \rangle = \sum_{m,l} e^{-i E_{m,l} t} |\psi_m \rangle_l \, \, {}_l\langle \psi_m | \psi(0) \rangle.
\end{equation} 
Here, $m$ is a mode index and ${}_l\langle \psi_m | \psi_{m'} \rangle_{l'} = \delta_{l-l'} \delta_{m-m'}$, where the Kronecker delta is defined such that $\delta_{0}=1$ and $\delta_i=0$ for $i\neq0$. For the initial condition $| e \rangle | 00 \rangle$ used in the manuscript, $\langle e | \langle 00  | \psi_m \rangle_l = d_{0,l}^m \delta_{l-1}$.

\section{Entanglement Entropy}

The form of the ansatz in (\ref{eq:ansatz}) indicates that the ground state can feature entanglement between the two cavity modes as well as entanglement between the two-level system (TLS) and the two cavity modes. The degree of entanglement between the TLS and the cavity is quantified by the bipartite entanglement entropy
\begin{equation}
S = - \operatorname{Tr}{\left[\hat{\rho}_{l, \textrm{TLS}} \operatorname{log}{\hat{\rho}_{l, \textrm{TLS}}} \right]}.
\end{equation}
Here the matrix logarithm is taken with base two and $\hat{\rho}_{l, \textrm{TLS}}$ is the reduced density matrix of the TLS, i.e., when the cavity modes are traced over,
\begin{equation}
\hat{\rho}_{l, \textrm{TLS}} = \operatorname{Tr_{\rm cav}}{\left[| \psi \rangle_l {}_l\langle \psi | \right]} = \sum_n |c_{n,l}|^2 | g \rangle \langle g | + |d_{n,l}|^2 | e \rangle \langle e |.
\end{equation}
The entanglement entropy is then
\begin{equation}
S = - \sum_n |c_{n,l}|^2 \operatorname{log}{\left[ \sum_{n'} |c_{n',l}|^2 \right]} - \sum_n |d_{n,l}|^2 \operatorname{log}{\left[ \sum_{n'} |d_{n',l}|^2 \right]}.
\end{equation}

\begin{figure}[!t]
\includegraphics[width=0.4\columnwidth]{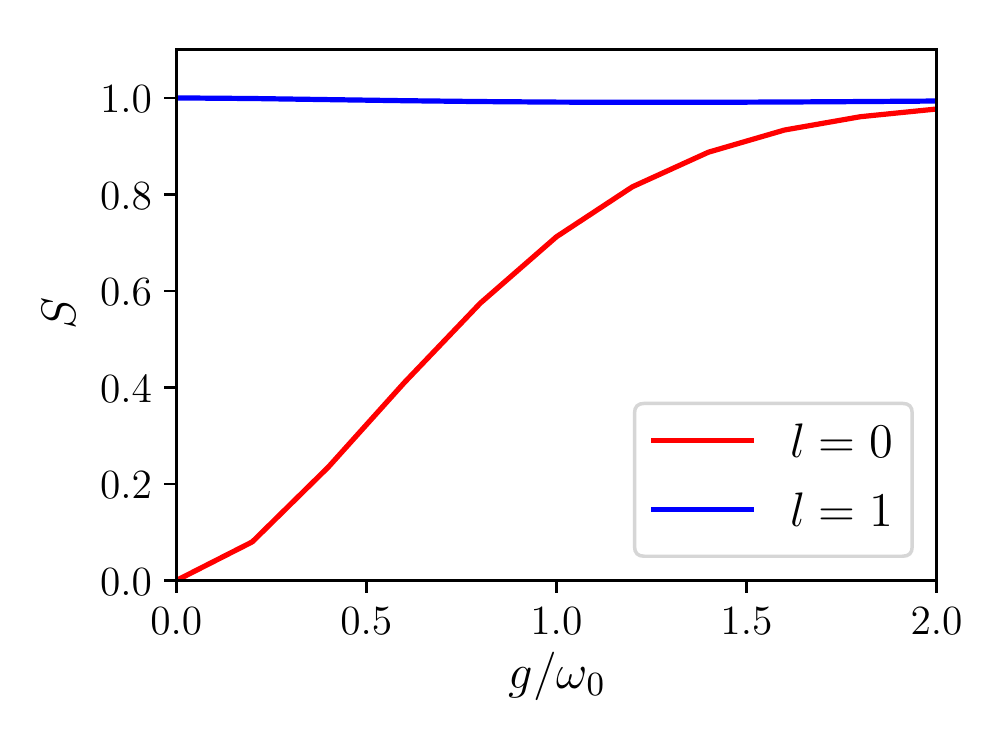}
\caption{\label{fig:entanglementEntropy} Bipartite entanglement entropy $S$ between the cavity and the TLS versus coupling strength $g/\omega_0$. The curves show the $l=0$ ground state (red) and the $l=1$ ground state (blue).}
\end{figure}

Figure \ref{fig:entanglementEntropy} shows the bipartite entanglement entropy between the TLS and the two cavity modes for the $l=0$ and $l=1$ ground states. When $g/\omega_0 \sim 0$ the $l=0$ ground state is $|g \rangle | 0 0 \rangle$ and does not feature entanglement. As $g/\omega_0$ increases the TLS becomes entangled with the two cavity modes and eventually becomes maximally entangled. The $l=1$ ground state has the form of a Jaynes-Cummings state with a single excitation when $g/\omega_0 \ll 1$ and is therefore maximally entangled. It remains maximally entangled as $g/\omega_0$ increases.

\section{Dynamics}

\begin{figure}[!t]
\includegraphics[width=\columnwidth]{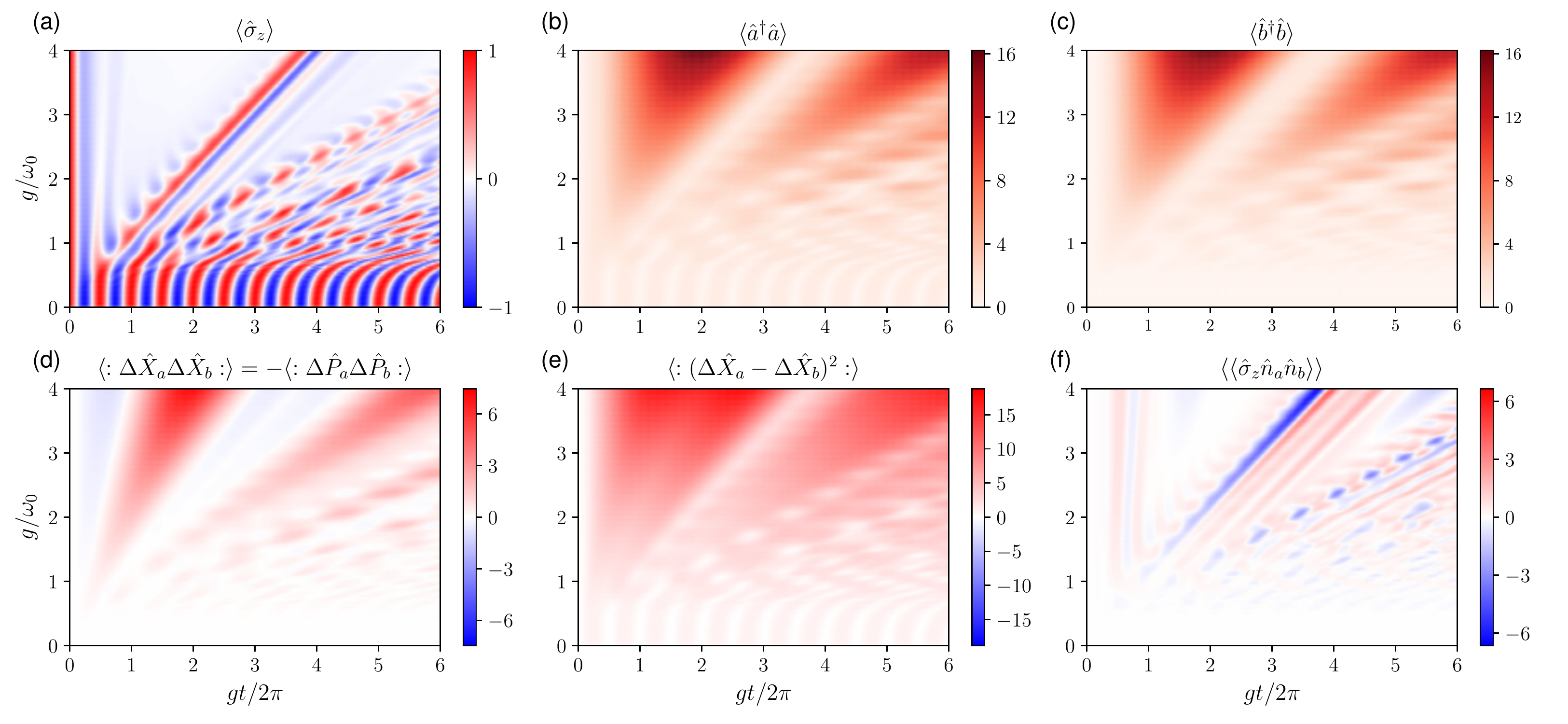}
\caption{\label{fig:dynEXTRA} Evolution of observables versus normalized time $gt/2\pi$ and coupling strength $g/\omega_0$ in chiral Rabi model ($\omega_0=\omega_c$) starting in state $| e\rangle |00\rangle$. The density plots show the (a) population $\langle \hat{\sigma}_z \rangle $, (b) photon number in mode $a$ $\langle \hat{a}^\dagger \hat{a} \rangle$, (c) photon number in mode $b$ $\langle \hat{b}^\dagger \hat{b} \rangle$, quadrature covariance $\langle : \Delta \hat{X}_a \Delta \hat{X}_b :\rangle$, normally ordered variance $\langle : (\Delta \hat{X}_a - \Delta \hat{X}_b)^2 :\rangle$, and  (d) the three-body cumulant $\langle \langle \hat{\sigma}_z \hat{n}_a \hat{n}_b \rangle \rangle =  \langle  \hat{\sigma}_z \hat{n}_a \hat{n}_b \rangle  - \langle \hat{\sigma}_z \hat{n}_a \rangle \langle \hat{n}_b \rangle  - \langle \hat{\sigma}_z\rangle \langle \hat{n}_a  \hat{n}_b \rangle  - \langle \hat{\sigma}_z \hat{n}_b \rangle \langle \hat{n}_a   \rangle + 2  \langle \hat{\sigma}_z \rangle \langle \hat{n}_b \rangle \langle \hat{n}_a   \rangle$.}
\end{figure}

One can also gain insight into the physics of this system by computing its dynamics. Here I consider the evolution of the initial state $|e\rangle| 0 0 \rangle$. The dynamics from this initial state can be efficiently computed by projecting $| e\rangle| 00 \rangle$ on the set of $l=1$ eigenstates and evolving them through time. Figure \ref{fig:dynEXTRA}~(a) shows the observable $\langle \hat{\sigma}_z \rangle$ versus time for different values of $g/\omega_0$. In the JC limit $g \ll \omega_0$  this exhibits well-known Rabi oscillations. Here, mode $b$ is decoupled from the dynamics. When $g \sim \omega_0$ the counter-rotating terms spoil the coherence of the Rabi oscillations. In the deep strong coupling limit $g> \omega_0$ the system dynamics descends into quasiperiodic collapses and revivals. In this limit the emitter rapidly decays into a state with $\langle \sigma_z  \rangle \sim 0$, while generating photons in modes $a$  and $b$ (see Fig.~\ref{fig:dynEXTRA}(b)-(c)). As shown in Fig.~\ref{fig:dynEXTRA}(d) the $X$ quadratures of these modes are highly correlated, but there is no squeezing (Fig.~\ref{fig:dynEXTRA}(e)). Once the system emits the maximum number of photons, which scales with $g^2$, the process reverses and the system starts absorbing the photons. This is similar to recurrences in the Rabi model \cite{Casanova2010PRL}. Once almost all the photons are absorbed the system is reexcited and $\langle \hat{\sigma}_z \rangle $ oscillates between values close to $1$ and $-1$. This oscillation is not a Rabi oscillation but results from full three-body interactions between photons in modes $a$ and $b$ and the TLS. This is evident from Fig.~\ref{fig:dynEXTRA}~(f) which  shows that the three-body cumulant $\langle \langle  \hat{\sigma}_z \hat{n}_a \hat{n}_b \rangle \rangle $ at the point  of this oscillation becomes non-zero.  The revivals are not complete as the system never fully reaches $\langle \hat{\sigma}_z \rangle =1$. Although the revivals occur periodically, their quality decays in time.


\section{Extension to a V-level scheme}

\begin{figure}[!t]
\includegraphics[width=0.5\columnwidth]{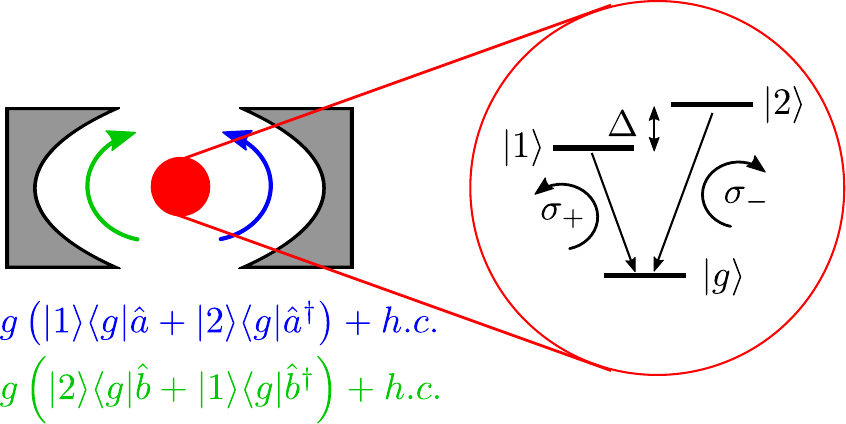}
\caption{\label{fig:Vschem} V-level scheme in a single cavity. An atom with three levels, $| g \rangle$, $|1 \rangle$, and $| 2 \rangle$, is coupled to a cavity with modes $a$ and $b$. Here the optical transition from $| 1 \rangle$ to $| g \rangle$ has a $\sigma_+$ polarization while the transition from $| 2 \rangle$ to $| g \rangle$ has a $\sigma_-$ polarization. These two polarizations are copolarized with the $a$ and $b$ modes respectively. The transitions couple to the copolarized modes via the rotating-wave terms and to the counter-polarized modes via counter-rotating terms.}
\end{figure}

The analysis in the manuscript can be extended to atoms with multiple levels. Here I consider an atom with a $V$-level scheme where the optical transitions have transition dipoles with opposite handedness circular polarization. As shown in Fig.~\ref{fig:Vschem}, the atom has levels $|g \rangle$, $|1 \rangle$, and $|2 \rangle$. As before $\hat{\mathbf E} = (\hat{a} \mathbf E_a + \hat{b} \mathbf E_b) +h.c.$, but now $\hat{\mathbf d}_{1} = \mathbf d_1 | g \rangle \langle 1 | + \mathbf d_1^* | 1 \rangle \langle g |$ and $\hat{\mathbf d}_{2} = \mathbf d_2 | g \rangle \langle 1 | + \mathbf d_2^* | 1 \rangle \langle g |$ with $\mathbf d_2 = \mathbf d_1^*$. Here, I take $\mathbf d_1$ to be copolarized with $\mathbf E_a$, while $\mathbf d_2$ is copolarized with $\mathbf E_b$. Within the dipole approximation, the Hamiltonian is
\begin{equation}
\hat{H}_V = \omega_c (\hat{a}^\dagger \hat{a} + \hat{b}^\dagger \hat{b}) + \omega_0 (|1 \rangle \langle 1| + |2 \rangle \langle 2| ) + \Delta | 2 \rangle \langle 2 | + g \left[ |1 \rangle \langle g| (\hat{a} + \hat{b}^\dagger)  + |2 \rangle \langle g| (\hat{b} + \hat{a}^\dagger)\right] + h.c..
\end{equation} 
Here the angular momentum, which is a conserved quantity and generates a $U(1)$ symmetry, is
\begin{equation}
\hat{L}_V = \hat{a}^\dagger \hat{a} + | 1 \rangle \langle 1| - (\hat{b}^\dagger \hat{b} + |2 \rangle \langle 2 |).
\end{equation} 
As before the conserved quantity is used to construct an ansatz for the eigenstates. The ansatz is
\begin{equation}
| \psi_V \rangle_l = \sum_{n=0}^\infty c_{n,l} |g \rangle | n+ l \rangle |n \rangle + d_{n,l} | 1 \rangle | n+l -1 , n \rangle + e_{n,l} | 2 \rangle | n+l+1, n \rangle. 
\end{equation} 

One can use the ansatz to derive a set of eigenrecurrence relations
\begin{equation}
\begin{split}
\omega_c (2n+l)c_{n,l} + g\left[d_{n,l} \sqrt{n+l} + d_{n+1,l} \sqrt{n+1} + e_{n-1} \sqrt{n} +  e_{n,l} \sqrt{n+l+1} \right] &= E_l c_{n,l}\\
\left[\omega_0 + \omega_c (2n+l-1) \right] d_{n,l} + g \left[ c_{n,l} \sqrt{n+l} + c_{n-1,l} \sqrt{n} \right] &= E_l d_{n,l}\\
\left[\omega_0 + \Delta + \omega_c (2n+l+1) \right] e_{n,l} + g \left[c_{n+1} \sqrt{n+1}  + c_{n,l} \sqrt{n+l+1}  \right] &= E_l e_{n,l},
\end{split}
\end{equation}
where I have taken $g$ to be real. The coefficients can be solved for by truncating at a sufficiently large Fock state $n$.

\section{Many-body ansatz}

In this section I present the eigenrecurrence equations for the many-body Hamiltonian. Starting from the Hamiltonian in Eq.~\ref{eq:MBHamiltonian} one uses the ansatz in Eq.~\ref{eq:MBansatz} and substitutes it into Schr\"{o}dinger's equation to obtain a set of eigenrecurrence equations.  Equations for each Fock state $n$ are obtained by projecting out the terms containing the ground state $| g\rangle$, 
\begin{equation}
\label{eq:MBbigEqns1}
\begin{split}
&\left[\omega_c (2n+l) - \frac{\omega_0}{2} \right] c_{n,l;j_1,j_2,\ldots,j_{n}}^{i_1,i_2,\ldots,i_{n+l}} + \frac{g}{\sqrt{n+l}} d_{n,l;j_1,j_2,\ldots,j_{n}}^{i_1,i_2,\ldots,i_{n+l-1}} \left[ \delta_{i_1} + \delta_{i_2} + \ldots + \delta_{i_{n+l}} \right]\\
&+ \frac{g}{\sqrt{n+1}}\left[ d_{n+1,l;0,j_1,j_2, \ldots,j_{n}}^{i_1,i_2,\ldots,i_{n+l}} + d_{n+1,l;j_1,0,j_2, \ldots,j_n}^{i_1,i_2,\ldots,i_{n+l}} + \ldots + d_{n+1,l;j_1,j_2, \ldots, j_n,0}^{i_1,i_2,\ldots,i_{n+l}} \right]\\
&-J \left[ c_{n,l;j_1,j_2,\ldots,j_n}^{i_1+1,i_2,\ldots,i_{n+l}} + c_{n,l;j_1,j_2,\ldots,j_n}^{i_1,i_2+1,\ldots,i_{n+l}} + \ldots + c_{n,l;j_1,j_2,\ldots,j_n}^{i_1,i_2,\ldots,i_{n+l}+1} \right]
-J \left[ c_{n,l;j_1,j_2,\ldots,j_n}^{i_1-1,i_2,\ldots,i_{n+l}} + c_{n,l;j_1,j_2,\ldots,j_n}^{i_1,i_2-1,\ldots,i_{n+l}} + \ldots + c_{n,l;j_1,j_2,\ldots,j_n}^{i_1,i_2,\ldots,i_{n+l}-1} \right]\\
&-J \left[ c_{n,l;j_1+1,j_2,\ldots,j_n}^{i_1,i_2,\ldots,i_{n+l}} + c_{n,l;j_1,j_2+1,\ldots,j_n}^{i_1,i_2,\ldots,i_{n+l}} + \ldots + c_{n,l;j_1,j_2,\ldots,j_n+1}^{i_1,i_2,\ldots,i_{n+l}} \right]\\
&-J \left[ c_{n,l;j_1-1,j_2,\ldots,j_n}^{i_1,i_2,\ldots,i_{n+l}} + c_{n,l;j_1,j_2-1,\ldots,j_n}^{i_1,i_2,\ldots,i_{n+l}} + \ldots + c_{n,l;j_1,j_2,\ldots,j_n-1}^{i_1,i_2,\ldots,i_{n+l}-1} \right]
 = E_l c_{n,l;j_1,j_2,\ldots,j_{n}}^{i_1,i_2,\ldots,i_{n+l}}. 
\end{split}
\end{equation}
The second set of equations are obtained by projecting out the terms containing the excited state $| e \rangle$,
\begin{equation}
\label{eq:MBbigEqns2}
\begin{split}
&\left[\omega_c (2n+l-1) + \frac{\omega_0}{2} \right] d_{n,l;j_1,j_2,\ldots,j_{n}}^{i_1,i_2,\ldots,i_{n+l-1}} + \frac{g}{\sqrt{n-1}} \left[ c_{n-1,l;j_2,j_3,\ldots,j_{n-1}}^{i_1,i_2,\ldots,i_{n+l-1}} \delta_{j_1} + c_{n-1,l;j_1,j_3,\ldots,j_{n-1}}^{i_1,i_2,\ldots,i_{n+l-1}} \delta_{j_2} \right.\\
&\left. + \ldots + c_{n-1,l;j_1,j_2,\ldots,j_{n-2}}^{i_1,i_2,\ldots,i_{n+l-1}} \delta_{j_{n-1}} \right] + \frac{g}{\sqrt{n+l}} \left[c_{n,l;j_1,j_2,\ldots,j_{n}}^{0,i_1,\ldots,i_{n+l-1}} + c_{n,l;j_1,j_2,\ldots,j_{n}}^{i_1,0,i_2,\ldots,i_{n+l-1}} + \ldots + c_{n,l;j_1,j_2,\ldots,j_{n}}^{i_1,i_2,\ldots,i_{n+l-1},0} \right]\\
&-J \left[ d_{n,l;j_1,j_2,\ldots,j_n}^{i_1+1,i_2,\ldots,i_{n+l-1}} + d_{n,l;j_1,j_2,\ldots,j_n}^{i_1,i_2+1,\ldots,i_{n+l-1}} + \ldots + d_{n,l;j_1,j_2,\ldots,j_n}^{i_1,i_2,\ldots,i_{n+l-1}+1} \right]\\
&-J \left[ d_{n,l;j_1,j_2,\ldots,j_n}^{i_1-1,i_2,\ldots,i_{n+l-1}} + d_{n,l;j_1,j_2,\ldots,j_n}^{i_1,i_2-1,\ldots,i_{n+l-1}} + \ldots + d_{n,l;j_1,j_2,\ldots,j_n}^{i_1,i_2,\ldots,i_{n+l-1}-1} \right]\\
&-J \left[ d_{n,l;j_1+1,j_2,\ldots,j_n}^{i_1,i_2,\ldots,i_{n+l-1}} + d_{n,l;j_1,j_2+1,\ldots,j_n}^{i_1,i_2,\ldots,i_{n+l-1}} + \ldots + d_{n,l;j_1,j_2,\ldots,j_n+1}^{i_1,i_2,\ldots,i_{n+l-1}} \right]\\
&-J \left[ d_{n,l;j_1-1,j_2,\ldots,j_n}^{i_1,i_2,\ldots,i_{n+l-1}} + d_{n,l;j_1,j_2-1,\ldots,j_n}^{i_1,i_2,\ldots,i_{n+l-1}} + \ldots + d_{n,l;j_1,j_2,\ldots,j_n-1}^{i_1,i_2,\ldots,i_{n+l-1}-1} \right] = E_l d_{n,l;j_1,j_2,\ldots,j_{n}}^{i_1,i_2,\ldots,i_{n+l-1}}.
\end{split}
\end{equation}
As in the main text, here the indices $i_k$ and $j_k$ run over lattice sites $0,1,2,\ldots,L-1$ for $k \geq 1$.

Diagonalizing the above equations becomes impractical by truncating at some large $n$ as the size of the Hilbert space grows exponentially as $L^{2n+l}$. The equations for $l=0$ and $l=1$ can be diagonalized for small systems $L \lesssim 30$ when keeping terms $n \leq 2$. These provide a good approximation for the $l=0$ and $l=1$ ground states which are dominated by $n=0$ and $n=1$ coefficients. When truncating at $n=2$, for $l=0$, one obtains the coupled set of equations ($d_{0,0}=0$) which are diagonalized to obtain the results shown in Fig.~\ref{fig:MBGS} of the main text. An equivalent set of equations can be obtained for $l=1$.

\section{Phase Transition of the many-mode system within the RWA}

The $l=1$ ground state of the many-mode system, whose energy and observables are shown in Figure~\ref{fig:MBGS}, undergoes a phase transition at a critical value of $g$. The behaviour of this phase transition can be described within the rotating wave approximation and is due to the semi-infinite geometry of the coupled-cavity waveguide array. In a finite cavity array (without periodic boundaries) with $L$ cavities in the absense of the TLS, the Hamiltonian is $\hat{H}_{\rm CCW} = \sum_{i=1}^{L} \omega_c \hat{a}_i^\dagger \hat{a}_i -J \sum_{i=1}^{L-1} \hat{a}_i \hat{a}_{i+1}^\dagger + h.c.$, and can be diagonalized by using the $k$-space operators with 
\begin{equation}
\begin{split}
\hat{a}_k &= \frac{1}{\sqrt{L}}\sum_{j=1}^L \hat{a}_j \sin{\left[ \frac{\pi k}{L+1} j \right]},\\
\hat{a}_j &= \frac{1}{\sqrt{L}}\sum_{k=1}^L \hat{a}_k \sin{\left[ \frac{\pi k}{L+1} j \right]},
\end{split}
\end{equation}
which gives $\hat{H}_{\rm CCW} = \sum_{k=1}^L \left[ \omega_0 - 2J \cos{\left( \frac{\pi k}{L+1} \right)}  \right] \hat{a}_k^\dagger \hat{a}_k$. The TLS is then introduced into the first cavity, which, within the RWA, leads to the interaction $g(\hat{\sigma}_+ \hat{a}_{i=1} + h.c.)$. The $l=1$ ground state then corresponds to the single-excitation eigenstate. This eigenstate can be described using the ansatz
\begin{equation}
|\phi \rangle = \sum_{k=1}^L \alpha_k \hat{a}^\dagger_k | 0, g \rangle + b \, \sigma_{+} | 0, g \rangle,
\end{equation}
where $| 0 , g\rangle$ corresponds to the TLS being in the ground state and all the cavities being in the vacuum state. Substituting $| \phi \rangle$ into Schr\"{o}dinger's equation leads to the equations
\begin{equation}
\begin{split}
\frac{g b}{\sqrt{L}} \sin{\left( \frac{\pi k}{L+1}  \right)} &= \left[ E - \omega_0 + 2J \cos{\left( \frac{\pi k}{L+1} \right)} \right] \alpha_k\\
\frac{g}{\sqrt{L}} \sum_k \alpha_k  \sin{\left( \frac{\pi k}{L+1}\right)}   &= (E - \omega_0) b.
\end{split}
\end{equation}
Solving for $E$ by taking the continuum limit in $k$, transforming the sum to and integral, and looking for solutions with $|E - \omega_0| >  2J$, yields
\begin{equation}
E_\pm = \frac{\omega_0}{2} \pm \frac{g^2/J}{\sqrt{g^2/J^2 - 1}}.
\end{equation}
At $g=\sqrt{2} J$, $|E - \omega_0|=2J $ and thus the expression is only valid for $g \geq \sqrt{2} J$. At $g=\sqrt{2} J$ bound state enters the continuum. This point corresponds to the boundary between regions I and II in Figure \ref{fig:MBGS}(a) within the RWA.

\section{Matrix-Product States}

Matrix product states (MPS) are used here to compute the dynamics of the many-body Hamiltonian \ref{eq:MBHamiltonian}. Open source software (open MPS) \cite{Wall2012NJP, Jaschke2018CompPhysComm} is used for the computations. In order to to use this code, the Hamiltonian is recast using hardcore Bosons
\begin{equation}
\label{eq:MBHamHC}
\begin{split}
\hat{{\cal H}}_{HC} &= \omega_0\, \hat{c}^\dagger \hat{c} +\omega_c \sum_{i=0}^{L-1} \hat{a}_i^\dagger \hat{a}_i + \hat{b}_i^\dagger \hat{b}_i + g \, \hat{c}^\dagger (\hat{a}_0 + \hat{b}_0^\dagger ) + h.c. - J \sum_{i=0}^{L-2} \left(\hat{a}_i^\dagger \hat{a}_{i+1} + \hat{b}_i^\dagger \hat{b}_{i+1} \right) + h.c. + U \hat{c}^\dagger \hat{c}^\dagger \hat{c} \, \hat{c}.
\end{split}
\end{equation}
Here the spin operators are replaced with the bosonic operators with $[\hat{c}, \hat{c}^\dagger] = 1$ and a nonlinear term $U \hat{c}^\dagger \hat{c}^\dagger \hat{c} \, \hat{c}$ is introduced. In the limit $U \rightarrow \infty$ the bosonic mode $c$ can only contain at most one excitation and thus behaves as a spin-1 TLS. The dynamics of this Hamiltonian are therefore equivalent to that of Eq.~(\ref{eq:MBHamiltonian}). The conserved angular momentum now becomes $\hat{L}_{HC} = \hat{c}^\dagger \hat{c} + \hat{a}^\dagger \hat{a} - \hat{b}^\dagger \hat{b}$. Note that the energies are renormalized by $+\omega_0/2$. 

The MPS ansatz is used to represent a pure state of a system with $L$ sites and open boundary conditions as 
\begin{equation}
| \psi \rangle = \sum_{n_1, n_2, \ldots, n_L} A^{n_1} A^{n_2} \ldots A^{n_L} | n_1 n_2 \ldots n_L \rangle,
\end{equation}
where the $A^{n_j}$ are matrices and the $n_j$ range over the number of bosonic excitations. For sufficiently large matrices and $n_j$ the representation is exact, but is exponentially large. If the matrices $A^{n_j}$ are scalars the ansatz can only represent product states. Matrices are required to represent entanglement with larger matrices being able to represent more entanglement. There are therefore two main convergence parameters that are checked here: the local bond-dimension $\chi$, which controls the amount of entanglement allowed in the system, and the number of Bosonic excitations $\nu$  allowed at each site, i.e., in modes $a$, $b$, and $c$ combined. 

\section{Comparsion of MPS and Ansatz}

\begin{figure}[!ht]
\includegraphics[width=\columnwidth]{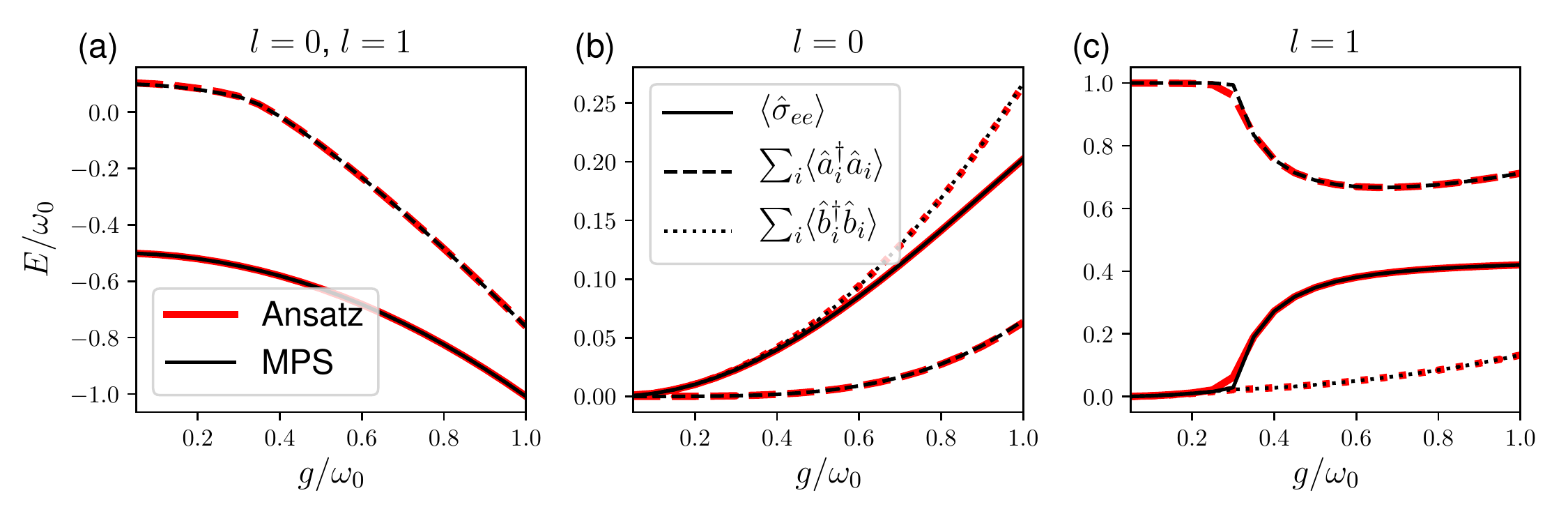}
\caption{\label{fig:MPSvsAnsatzEigenstates} Comparison of eigenstates computed using MPS (black) and the Ansatz (red) showing (a) the energy, and the observables for the (b) $l=0$ and  (c) $l=1$ eigenstates  with $J=0.2\omega_0$ and $\omega_0=\omega_c$.  For the MPS calculations $L=100$, the maximum bond dimensions is $\chi=10$ and maximum number of bosons is $\nu=5$. For the ansatz $L=20$.}
\end{figure} 

Figure \ref{fig:MPSvsAnsatzEigenstates} compares the eigenstate calculations shown in Fig.~\ref{fig:MBGS} of the main text with MPS calculations. In the MPS calculations a much larger computation domain of $L=100$ was used. Both the observables and the energies show excellent agreement. For $l=1$, in both the ansatz and MPS calculations the photons in the $a$ mode did not decay to zero at the computation boundaries for $g \leq 0.3 \omega_0$. The eigenstates are thus influenced by the boundary conditions and these points are therefore omitted from the plots.

\begin{figure}[!ht]
\includegraphics[width=\columnwidth]{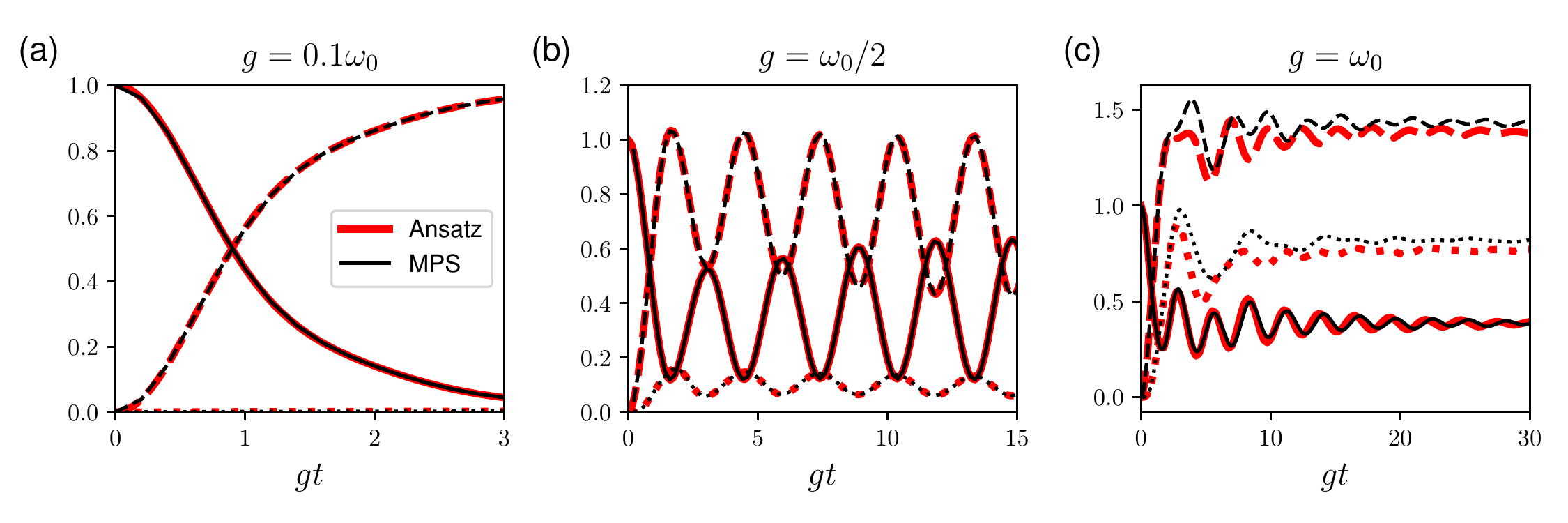}
\caption{\label{fig:MPSvsAnsatz} Comparison of quench dynamics computed using MPS (black) and the Ansatz (red) for (a) $g=0.1\omega_0$, (b) $g=\omega_0/2$, and (c) $g=\omega_0$ with $J=0.2\omega_0$ and $\omega_0=\omega_c$. In all plots solid lines show $\langle \hat{\sigma}_{ee} \rangle$, dashed lines show $\sum_i \langle \hat{a}_i^\dagger \hat{a}_i \rangle$, and dotted lines show $\sum_i \langle \hat{b}_i^\dagger \hat{b}_i \rangle$. For the MPS calculations the maximum bond dimensions $\chi$ and boson number $\nu$ are (a) $\chi=10$ and $\nu=5$, (b) $\chi=20$ and $\nu=7$, and (c) $\chi=30$ and $\nu=7$.}
\end{figure}

Figure \ref{fig:MPSvsAnsatz} shows the comparison of observables versus time for the quench dynamics computed using the many-body ansatz in equation (\ref{eq:MBansatz}) truncated at $n \leq 2$ and an MPS ansatz. The parameters used here for the MPS simulations are the same as in Fig.~\ref{fig:MBDynamics} in the main text. The two computations show excellent agreement for $g=0.1 \omega_0$ and $g=\omega_0/2$, while the ansatz starts becoming quantitatively inaccurate at $g=\omega_0$. This is because for larger values of $g$, the number of photons in the system becomes large and cannot be captured by the ansatz when it is truncated to values $n \leq 2$.

\begin{figure}[!ht]
\includegraphics[width=\columnwidth]{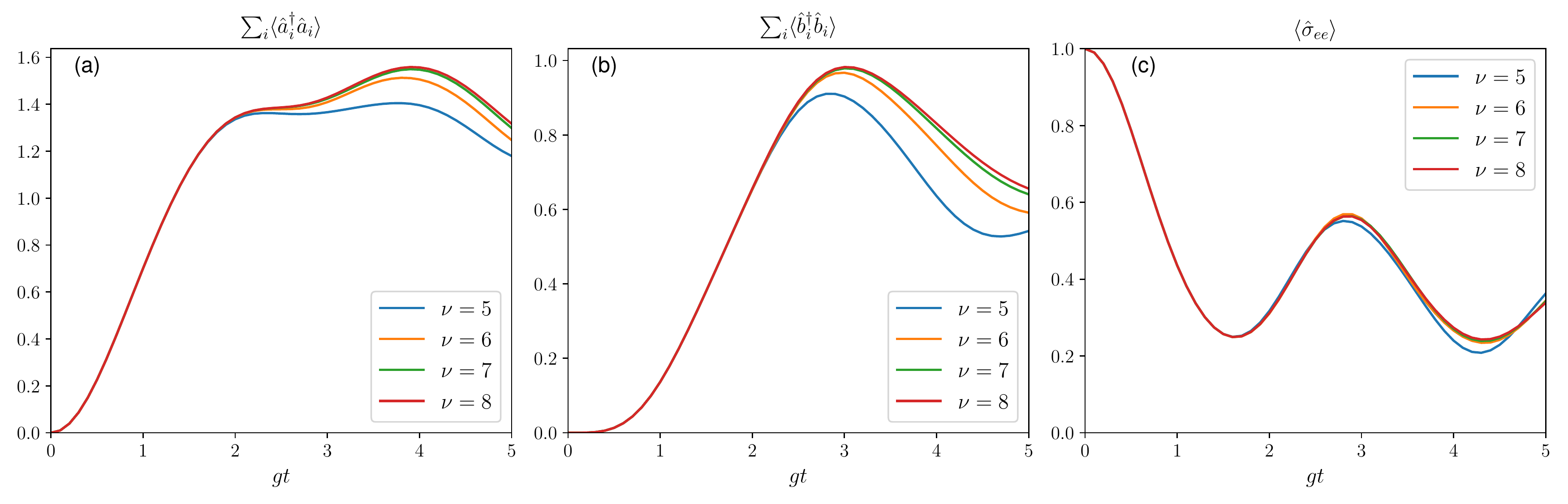}
\caption{\label{fig:MPSquenchBosonConv} Convergence of the many-body quench calculations for different maximum number of Bosonic excitations on each site $\nu$. The system starts in $|e \rangle | 0 0 \rangle$ and evolves in time. The simulation parameters are for $\omega_0=\omega_c=g$, and $J=0.2\omega_0$. The frames show (a) the total number of photons in mode $a$, $\sum_i \hat{a}^\dagger_i \hat{a}_i$ (b) in mode $b$, $\sum_i \hat{b}^\dagger_i \hat{b}_i$, and (c) the population of the excited state $\langle \hat{\sigma}_{ee} \rangle$.}
\end{figure}

\begin{figure}[!ht]
\includegraphics[width=\columnwidth]{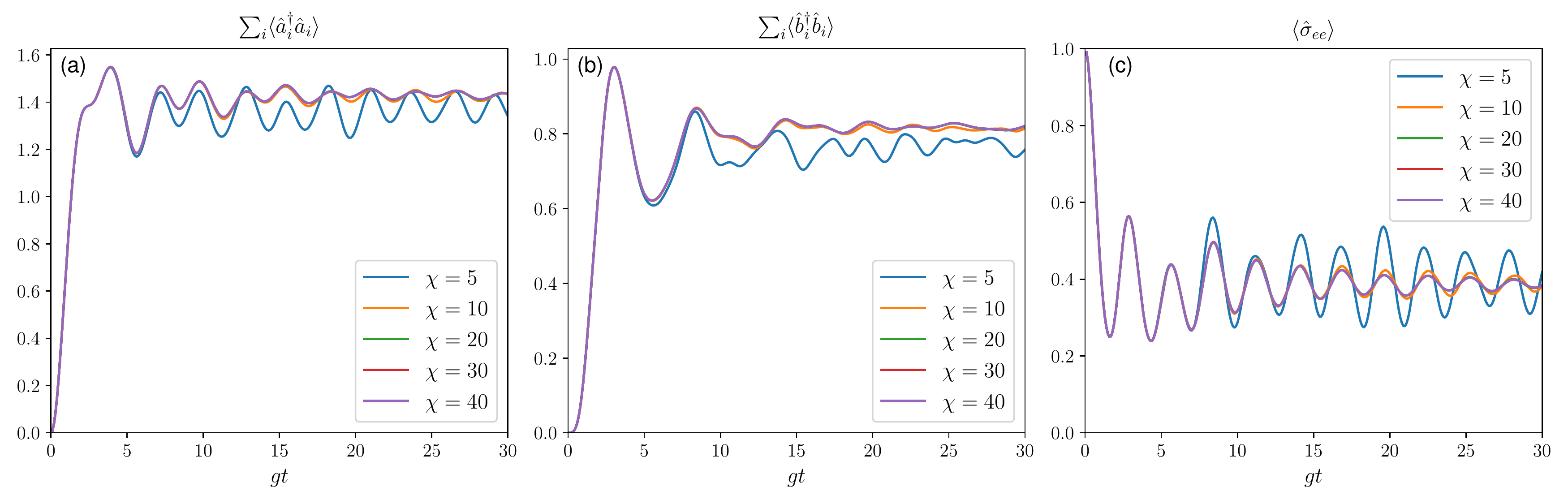}
\caption{\label{fig:MPSquenchBondConv} Convergence of the many-body quench calculations for different bond dimensions $\chi$ with $\nu=7$. The system starts in $|e \rangle | 0 0 \rangle$ and evolves in time. The simulation parameters are for $\omega_0=\omega_c=g$, and $J=0.2\omega_0$. The frames show (a) the total number of photons in mode $a$, $\sum_i \hat{a}^\dagger_i \hat{a}_i$ (b) in mode $b$, $\sum_i \hat{b}^\dagger_i \hat{b}_i$, and (c) the population of the excited state $\langle \hat{\sigma}_{ee} \rangle$.}
\end{figure} 

\section{Convergence calculations for MPS}

Since there is a discrepancy between the results computed using MPS and the ansatz for $g=\omega_0=\omega_c$, a convergence analysis of the MPS calculations is performed. Figures \ref{fig:MPSquenchBosonConv} and \ref{fig:MPSquenchBondConv} show the convergence calculations for the quench dynamics with $g=\omega_0=\omega_c$ and $J=0.2\omega_0$ using MPS for the maximum number of bosons and the bond dimensions respectively. The evolution of the system here is such that the dynamics is most sensitive to the value of $\nu$ at short times, while being sensitive to the value of $\chi$ at larger times. This is because near the beginning of the calculations, the emitter produces many photons at site $i=0$ and therefore this part of the evolution determines the required value of $\nu$. On the other hand as the simulation progresses the amount of entanglement across the sites grows and, at larger times, larger bond dimensions $\chi$ are required to capture the entanglement in the quantum state. This is rather fortunate, because the convergence of $\nu$ and $\chi$ can then be done independently. One chooses a smaller value of $\chi=20$ and runs the simulation for different values of $\nu$ for short times to check this parameter for convergence. This is shown in Fig.~\ref{fig:MPSquenchBosonConv}. These computations indicate that $\nu=7$ is adequate to obtain reasonable convergence for understanding the dynamics of the system. Once the value of $\nu=7$ is set, the convergence calculations for $\chi$ can be carried out. This is shown in Fig.~\ref{fig:MPSquenchBondConv} where the dynamics are simulated to $gt=30$. Clearly a bond dimension of $\chi=5$ is insufficient for obtaining even qualitatively accurate results. From these computations it appears that $\chi \geq 20$ is required for capturing the correct time-evolution of the observables. The convergence here is performed for $g=\omega_0$. For smaller $g$, fewer photons are generated and thus lower values of $\nu$ and $\chi$ can be used. This convergence study therefore gives an upper bound for the required values of $\nu$ and $\chi$.

\end{document}